\documentclass[multphys,vecphys,graphics]{svmult}

\usepackage{makeidx}     
\usepackage{graphicx}    
\usepackage{multicol}    

\makeindex             

\begin{document}

\title*{SN 1994W: evidence of explosive mass ejection 
a few years before explosion}
\titlerunning{SN 1994W: evidence of explosive mass ejection before explosion}
\author{Nikolai N. Chugai\inst{1} \and Robert J. Cumming\inst{2} \and Sergei I. Blinnikov\inst{3} \and
Peter Lundqvist\inst{2} \and Alexei V. Filippenko\inst{4}  \and Aaron J.
Barth\inst{4}   \and Angela Bragaglia\inst{5}   \and  Douglas C. Leonard\inst{6}  
\and Thomas Matheson\inst{7}   \and Jesper
Sollerman\inst{2}  }
\authorrunning{N. N. Chugai et al.}
\institute{Institute of Astronomy, RAS, Pyatnitskaya 48, 109017 Moscow,
 Russia.
\texttt{nchugai@inasan.rssi.ru}
\and Stockholm Observatory, AlbaNova University Center, SE-106~61
Stockholm, Sweden. \texttt{robert@astro.su.se, peter@astro.su.se}  \and ITEP, 117218 Moscow, Russia. \texttt{blinn@sai.msu.su} \and Department of Astronomy, University of California at Berkeley, Berkeley,
CA 94720-3411, USA. \and Osservatorio Astronomico di Bologna, via Ranzani 1, 40127 Bologna, 
Italy \and Department of Astronomy,
University of Massachusetts,
 710 North Pleasant Street,
Amherst, MA 01003-9305, USA. \and Harvard-Smithsonian Center for Astrophysics,
Mail Stop 20,
60 Garden Street,
Cambridge, MA 02138, USA.}
%
%
\maketitle

\begin{abstract}

We present and analyse spectra of the Type IIn supernova 1994W
obtained between 18 and 202 days after explosion.  During the first 100 days the line profiles are composed of three major
components: (i) narrow P Cygni lines with absorption minima at
$-700$ km s$^{-1}$; (ii) broad emission lines with blue velocity at
zero intensity $\sim 4000$ km s$^{-1}$; (iii) broad, smooth, extended
wings most apparent in H$\alpha$.  These components are identified
with the expanding circumstellar (CS) envelope (Sollerman, Cumming \&
Lundqvist 1998), shocked cool gas in the forward postshock region, and
multiple Thomson scattering in the CS envelope, respectively.  The
absence of broad P Cygni lines from the supernova (SN) is the result
of the formation of an optically thick, cool, dense shell at the
interface of the ejecta and the CS envelope.  Models of the SN
deceleration and Thomson scattering wings are used to recover the
Thomson optical depth of the CS envelope, $\tau_{\rm T}\geq 2.5$
during first month, its density ($n\sim 10^9$ cm$^{-3}$) and radial
extent, $\sim (4-5)\times 10^{15}$ cm.  The plateau-like SN light
curve, which we reproduce by a hydrodynamical model, is 
powered by a combination of internal energy leakage after the
explosion of an extended presupernova ($\sim 10^{15}$ cm) and
subsequent luminosity from circumstellar interaction.  We recover the
pre-explosion kinematics of the CS envelope and find it to be close to homologous
expansion with outmost velocity $\approx1100$ km s$^{-1}$ and a
kinematic age of $\sim 1.5$ yr.  The high mass
($\approx0.4~M_{\odot}$) and kinetic energy ($\approx 2\times10^{48}$
erg) of the CS envelope combined with small age strongly suggest that
the CS envelope was explosively ejected only a few years before the
SN explosion. 
\end{abstract}

  \begin{figure*}
  \centering
\includegraphics[height=12cm]{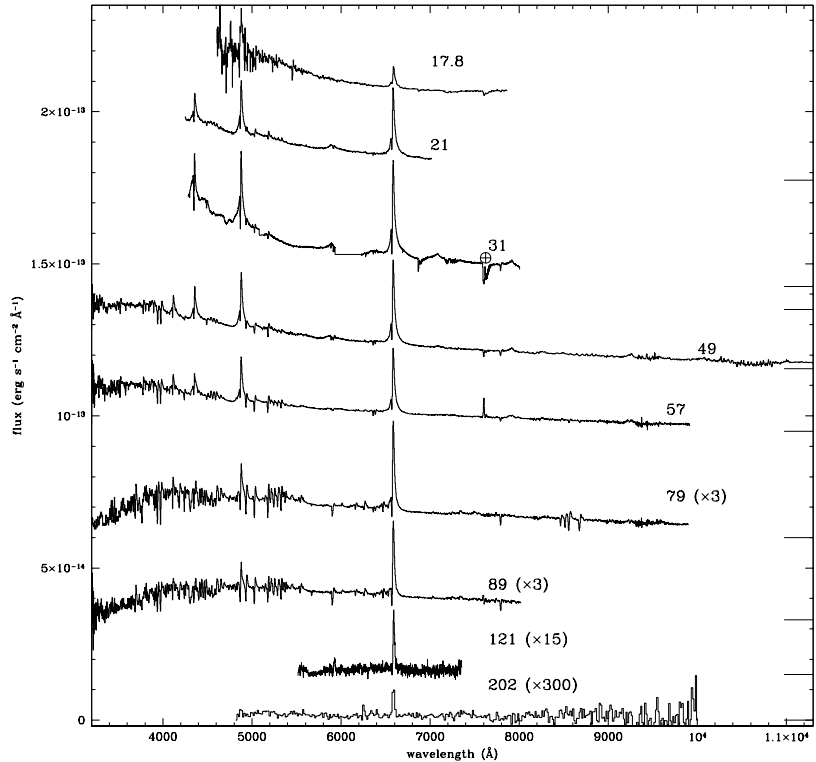}
  \caption[Spectra of SN 1994W: days 18-202]{Spectra of SN 1994W: days
18-202.  The spectra have been shifted vertically for clarity.  The
ticks on the right hand side mark the zero level for each.  The 
spectra from day 79 and later have been multiplied by a constant,
noted in brackets.  Note the change in relative intensity of the
Balmer lines, a sign of temperature evolution.}
  \label{f-stack}
  \end{figure*}

\section{Introduction}
\label{sec-intro}

Type IIn supernovae are believed to arise from massive stars exploding
into a dense circumstellar environment.  Often highly luminous, they
offer us an intriguing new window on the final stages of the progenitor's
evolution.

SN 1994W, discovered on 1994 July 29 in NGC 4041, was a luminous Type
IIn supernova whose light curve dropped dramatically at 110 days.  The
low luminosity after this point has been used to derive a very low
mass of nickel in the ejecta ($M_{\rm Ni}<$0.003 M$_{\odot}$;
\cite{SCL98}).  In a forthcoming paper \cite{C03}, we present and
analyse all the SN 1994W spectra and model the circumstellar
interaction.

\section{Observations}

The spectra were taken between 1994 July and 1995 February using BFOSC
on the BAO 1.5-m telescope, the IDS on the Isaac Newton Telescope,
ISIS on the William Herschel Telescope (La Palma), the Kast
spectrograph at Lick Observatory's Shane 3-m reflector, and the LDS at
the Nordic Optical Telescope on La Palma.

\section{Spectral evolution}

No broad ejecta absorption lines are seen (Figure \ref{f-stack}).
Together with high luminosity at maximum, this suggests an extended
progenitor interacting with a circumstellar (CS) envelope, and the
presence of a cool dense shell (CDS), which should form in SN~II with
extended envelopes \cite{FA77,G71}.

\begin{figure}
  \centering
\includegraphics[height=7cm]{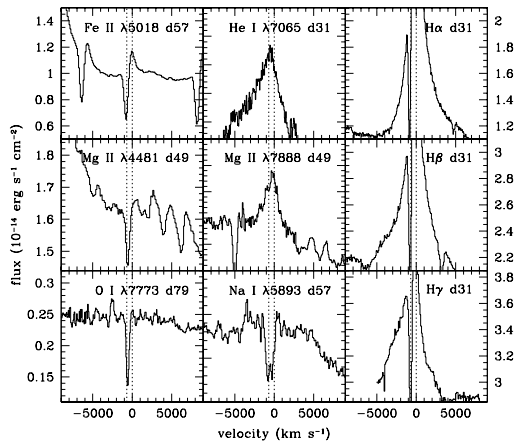}
  \caption[Line profiles.]{Selected line profiles.  The left hand
panel shows examples of narrow P Cygni profiles, the middle panel
shows broad lines (note the extended wings in Na~{\sc i} on day 57), and
the right hand panel shows the wings of the H~{\sc i} lines on day 31.  The dotted
vertical lines mark velocities of 0 and $-$700 km~s$^{-1}$.}
  \label{f-profiletypes}
  \end{figure}

The spectra show persistent narrow P~Cygni lines of H~{\sc i} with
broad bases (Figure \ref{f-profiletypes}).  Triangular profiles with $v_{\rm FWHM}\sim2500$
km~s$^{-1}$ are seen in He~{\sc i} and Mg~{\sc ii}.  Black body fits
to the continuum show that the temperature declines from 15000~K on
day 31 to 7200~K on day 89.  From day 121 onward, only narrow emission
lines are clearly seen.

\label{sec-profiles}

 \begin{figure}
  \centering
\includegraphics[height=7cm]{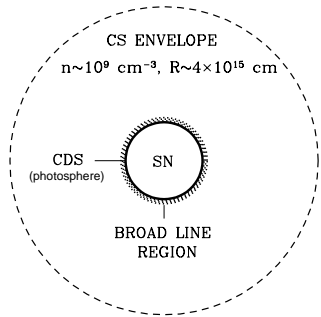}
  \caption []{A visualization of SN~1994W at around day 30.
The SN ejecta is bounded by an opaque cool dense shell
(CDS), which is responsible for the continuum radiation. The broad
line region is a narrow mixing layer attached to the CDS, made up of
Rayleigh-Taylor fragments of the CDS matter and possibly of shocked CS
clouds.  The SN ejecta expands into a dense CS envelope with Thomson
optical depth of  order unity.  The CS envelope is responsible
for both narrow lines and the extended Thomson wings seen in
H$\alpha$.
  }
  \label{f-cart}
  \end{figure}

We see broad lines with maximum velocity 4000 km~s$^{-1}$, which we
take to be the expansion velocity of the CDS.  We think the broad
lines come from Rayleigh-Taylor fingers from the interface of the
radiative forward shock and the CS envelope (Figure \ref{f-cart}).
The persistent, narrow P Cygni lines with maximum blue velocity 1100
km~s$^{-1}$ indicate the maximum velocity of the circumstellar
envelope, which is overrun by the ejecta at 110 days.  Broad wings on
H~{\sc i}, and nearly inverse Balmer decrement on day 31 point to a
high optical depth for Thomson scattering, from which we estimate that
the density in the CS envelope is as high as $\sim10^9$ cm$^{-3}$.

\section{Light curve models confirm energetics}

  \begin{figure}
  \centering
\includegraphics[height=7cm]{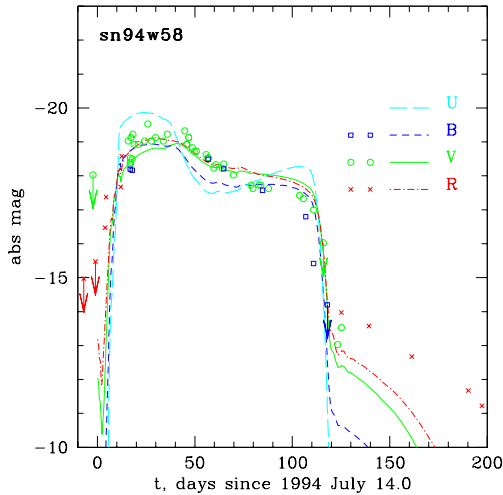}
  \caption[]{Light curves for a model of SN 1994W characterized by
$E=1.5\times10^{51}$ erg, $M_{\rm Ni}= 0.015$ M$_{\odot}$ and $M_{\rm ej}= 7$ M$_{\odot}$. }
  \label{lc58}
  \end{figure}

We model the broad band light curves using the multi-energy group
radiation hydrodynamic code {\sc STELLA}.  The best fit model is for a
$1.5\times10^{51}$ erg explosion, $M_{\rm Ni}=$0.015 M$_{\odot}$ and
$M_{\rm ej}=7$ M$_{\odot}$ ejecta, with circumstellar envelope
extending out to $R_{\rm out}=4.5\times10^{15}$ cm (Figure
\ref{lc58}).  A similar model, with the same bulk mass but no CS
envelope, fails to reproduce the bright plateau up to 110 days.

\section{H$\alpha$ profile model reveals kinematics}
\label{sec-kinem}

We have also modelled the evolution of the H$\alpha$ line profile.
The CS envelope is ionized by radiation from the forward shock,
leading to line emission from recombination and collisional
excitation.  Our best fit is for a model with free expansion in the CS
envelope and a boundary velocity of 1100 km~s$^{-1}$.  This suggests a
mass ejection $\sim$1.5 yr before explosion.

Some SN IIn seem to be due to interaction with a superwind.  SN 1994W,
on the other hand, like SN 1995G \cite{CD03}, seems instead to have
been preceded by a violent ejection event shortly before explosion,
perhaps due to a Ne flash \cite{CD03,WW86}.

%
%

%
%



\printindex
\end{document}